# An Agent-based Modelling Framework for Driving Policy Learning in Connected and Autonomous Vehicles

Varuna De Silva, Xiongzhao Wang, Deniz Aladagli, Ahmet Kondoz, Erhan Ekmekcioglu
Institute for Digital Technologies, Loughborough University, London, UK
Email: varunax@gmail.com, A.D.Aladagli@lboro.ac.uk, A.Kondoz@lboro.ac.uk, and E.Ekmekcioglu@lboro.ac.uk

*Abstract*—Due to the complexity of the natural world, a programmer cannot foresee all possible situations, a connected and autonomous vehicle (CAV) will face during its operation, and hence, CAVs will need to learn to make decisions autonomously. Due to the sensing of its surroundings and information exchanged with other vehicles and road infrastructure, a CAV will have access to large amounts of useful data. While different control algorithms have been proposed for CAVs, the benefits brought about by connectedness of autonomous vehicles to other vehicles and to the infrastructure, and its implications on policy learning has not been investigated in literature. This paper investigates a data driven driving policy learning framework through an agent-based modelling approaches. The contributions of the paper are two-fold. A dynamic programming framework is proposed for in-vehicle policy learning with and without connectivity to neighboring vehicles. The simulation results indicate that while a CAV can learn to make autonomous decisions, vehicle-to-vehicle (V2V) communication of information improves this capability. Furthermore, to overcome the limitations of sensing in a CAV, the paper proposes a novel concept for infrastructure-led policy learning and communication with autonomous vehicles. In infrastructure-led policy learning, road-side infrastructure senses and captures successful vehicle maneuvers and learns an optimal policy from those temporal sequences, and when a vehicle approaches the road-side unit, the policy is communicated to the CAV. Deep-imitation learning methodology is proposed to develop such an infrastructure-led policy learning framework.

I. INTRODUCTION

The emergence of connected and autonomous vehicles (CAVs) marks a new phase of innovation in automotive and transportation industries since the development of personal automobiles [1]. Once technologically matured, CAVs would be able to move autonomously without the aid of a human driver and be able to communicate with other vehicles and traffic infrastructure. The promised benefits of CAVs are numerous, such as, reduced congestion, better fuel efficiency, reduced environmental pollution, reduced number of traffic related casualties and increased personal independence. The continuing evolution of sensor technologies, high speed communication infrastructure, machine learning and artificial intelligence will eventually bring us the above benefits.

A CAV is composed of four major technological components. The first is the perception system, which is responsible for sensing the environment to understand its surroundings. The second component is the localization and mapping system that enables the vehicle to know its current location. The third component is responsible for the driving policy. The driving policy refers to the decision making capability of a CAV under various situations, such as negotiating at roundabouts, giving way to vehicles and pedestrians, and overtaking vehicles. Finally, the CAVs will be connected. It is expected that the CAVs will be connected to the surrounding vehicles: vehicle to vehicle connectivity (V2V), to the infrastructure: Vehicle to Infrastructure (V2I) and to anything else such as the internet: Vehicle to Anything (V2X), through wireless communications links [2]. While there are many challenges still to be addressed for high speed wireless connectivity for vehicular applications [3], the IEEE802.11p Wireless Access in Vehicular Environments (Wave) is considered the most relevant standard that currently caters to the requirements of such applications [4].

The connectedness of the CAV's can be useful for many important functions related to intelligent mobility. One important use case is to exchange sensor data between vehicles for improved perception of the surroundings [5], which enables to reduce the accidents. Connected vehicles also enables centralized traffic control to ease congestion in smart city applications [6]. For example in Compass4D, a pilot project funded by the EU, it was demonstrated that 15% reductions in fuel efficiency can be achieved through V2I communication to control the flow of traffic [7]. Another interesting application of connectivity is in "vehicle platooning", where a group of vehicles with common interests maintain a small and constant distance to each other [8]. Connectivity is also a key enabler of intelligent mobility applications such as ubiquitous taxi services. Furthermore, various infotainment applications such as music/video streaming or in-car WiFi connectivity, are expected to be delivered to the CAVs through high speed wireless internet connectivity [4]. All the above examples lead to a significant amount of data that are been collected and shared between CAVs. This leads to the question, whether CAV could learn interesting patterns from these data? Could a CAV autonomously evolve to make better decisions over time? Data driven approaches are emerging within the scope of CAV research to utilize this data in effective ways. In [9], authors propose a data-driven control algorithm derivation for connected vehicle to account for unknown system dynamics such as human factors.

In this paper we investigate opportunities that arise from the proliferation of data within a CAV. In particular, this paper focuses on optimal policy learning in autonomous vehicles over its life time. For example, how could it make better decisions to reach its destination while avoiding crashes with other objects or road side infrastructure? We consider a use case where a CAV is able to make driving decisions based on information shared by other CAVs. The proposed framework is based on agent based learning paradigm where a CAV is modelled as a rational agent that tries to maximize its utility. Furthermore, an infrastructure-led policy learning and communicating framework is proposed based on deep-imitation learning.

The rest of this paper is organized as follows: Section II provides an overview of the agent based learning paradigm; Section III describes the experimental setup and the results; and Section IV concludes the paper with reference to some future work.

## II. RELATED WORK

In this section we will discuss work related to driving policy learning in connected autonomous vehicles. This section is organized as follows: Section A describes work related to policy learning in CAVs, including applications of Vehicle-to-Infrastructure (V2I) communication. Section B describes the foundations of imitation learning and its applications in driverless vehicle technology.

### A. Policy Learning in CAVs

Policy learning is the process by which an autonomous vehicle comes to learn which action to take given a particular situation. The situation or the state in which the CAV is operating is identified through sensing the environment and understanding it.

Different types of control problems were formulated for CAVs considering the cases of range-limited V2V communication and input saturation to minimize the errors of distance and velocity and to optimize the fuel usage [9]. In [9] authors employed an adaptive dynamic programming technique, to derive the optimal controllers without relying on the knowledge of system dynamics. The effectiveness of the proposed approaches in [9] was demonstrated via the online learning control of the connected vehicles in platform for traffic microsimulation.

Reinforcement learning was utilized for longitudinal control of autonomous vehicles [10]. A deep Reinforcement learning algorithm based on neural network-based Q-function approximation was proposed for ramp merging process for autonomous vehicles [11]. The ramp merging process involves interactions with other vehicles whose behaviors dynamic and varied and which influences the actions of the merging vehicles. Similarly, reinforcement learning based controllers were proposed for autonomous vehicles for tasks such as lane changing [12] and overtaking decisions[13].

While policy learning in CAV is an emerging area of research, agent-based models are utilized for various control functions in vehicle control. A learning based cruise control algorithm was proposed in [14].

### B. Applications of Deep Imitation Learning

Imitation learning is the process of learning by imitating certain set of actions. Imitation learning from human drivers for the task of lane keeping assistance in highway and country roads using grayscale images from a single front view camera was proposed in [15]. The employed method in [15] utilizes convolutional neural networks (CNN) to analyze the images to develop an appropriate policy to adhere to the lanes.

Deep imitation learning was proposed as a method to model the defensive behavior of football players [16]. In [16], Multiview video based player tracking data are collected and the team behavior is learnt. For this purpose, a recurrent neural network, specifically a Long-Short Term Memory (LSTM) network is trained with sequences of player tracking data.

### C. Contributions of this Paper

While different aspects of control systems have been developed for CAVs, the advantages of connectedness of vehicles and its implications for policy learning has not been investigated in literature.

The contributions of this paper are twofold: firstly, we propose a dynamic programming approach for policy learning in an autonomous vehicle and compare it with a situation where the vehicles are connected to each other to share sensor information. Secondly, we propose a framework for policy learning at road-side units, and for policies to be communicated to approaching vehicles. This way road-side units at which many examples of successful vehicle maneuverers can be observed, could be utilized for the safety of CAVs.

## III. AGENT BASED MODELLING FRAMEWORK FOR CAVs

In this section we will discuss the agent based learning framework that will be used for algorithm development in this paper.

### A. Foundations of Reinforcement Learning

In the recent past, reinforcement learning has been utilized in attractive research projects. DeepMind, a modern technique research organization hosted in London, which concentrates on data mining and machine learning, has made landmark progress on reinforcement learning research. The world champion in 'Go', Lee Sedol was beaten by the deep mind developed program called AlphaGo [17]. Fig. 1 illustrates the fundamental process of reinforcement learning. Actions taken in the current environment would transfer the agent to a different state, and a reward would be given based on how good the action was. A utility function that assigns a different reward value for each possible actions in the same state, could be used to evaluate the best action in that particular state. This interaction scenario gives a general idea of reinforcement learning and builds the foundation cooperating with Markov Decision Processes (MDP) and Bellman equations [18].

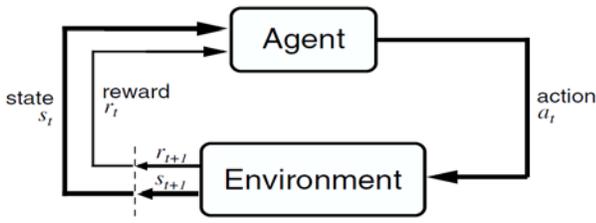

Fig. 1. The reinforcement learning interaction between the agent and its environment.

Markov decision processes are fundamental to the development of active reinforcement learning algorithms. The state transition in RL can be described as a Markov process, where the future states are dependent on the current state only and are independent of the states in the past. An active reinforcement learning approach known as Q-value iteration is used to find the optimal reward function and the transition model. Q-value iteration uses direct rewards value as samples, and update the corresponding Q-value by a proportional summation with the initial value, which depends on learning rate $\alpha$ as given below:

$$sample = R(s, a, s') + \gamma \cdot \max_{a'} Q(s', a'). \quad (1)$$

$$Q(s, a) \leftarrow (1 - \alpha) \cdot Q(s, a) + \alpha \cdot sample. \quad (2)$$

Where $R(s,a,s')$ is the immediate reward value given when an action $a$ is taken at state $s$, which moves the agent to state $s'$. $Q(x,y)$ in (1) and (2) refers to the expected future reward if action y is taken at state x. The agent acts in a rational manner to maximize its discounted future reward. $\gamma$ is the discount factor. A dynamic programming approach is used to solve the above equations in an iterative manner.

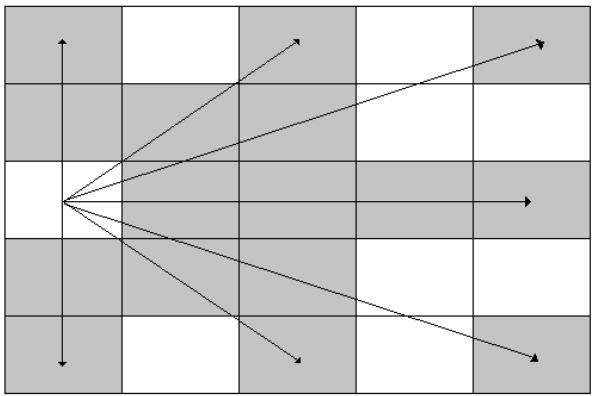

Fig. 2. The structure of the environmental scanner on the simulated CAV.

TABLE I.   IMMEDIATE REWARDS FUNCTION

| Condition (State/Action) | Value |
|---|---|
| Alive or Goal | 0.1 |
| Shift to right or left | -0.1 |
| Crash or Bump | -10 |
| Speed less than limit and alive or goal | speed/10 |
| Speed less than limit with bump or crash | - speed |
| Speed over limit | - 2×speed |

### B. Micro-simulation Framework for CAV Policy Learning

The CAV is simulated as a car that acquires sensor data of its surroundings and exchanges speed information with other vehicles in its surroundings through V2V communication. Each vehicle is simulated as a point on a 2D grid and ignores any spatial parameters such as the size of the vehicle. Fig. 2 illustrates the structure of the environmental scanner implemented on the CAV, which simulates a 7 directional scanner.

At a given state, the agent can take nine (3x3) actions, which control the movement direction and speed. They are the three direction actions: maintain current direction, which mean no change in the status of vehicle, move left and move right and the three speed actions: maintain, increase and reduce the speed of agent. Other decisions, such as making turns at intersections or stop at target point, will not be considered in the current study. The immediate reward function for the agent is summarized in Table I.

The prototype environment simulates a straight road that is 66 units long with two lanes of 1 unit width. The movement of each agent is simulated as a maze walk, where they move number of units at each time step according to the current speed. There would be a number of other randomly located vehicles, acting as obstacles for the intelligent agent, trying to move to the same destination. In this experiment we consider only one intelligent agent that adapts its decisions over time and the rest of the vehicles will move at a constant speed over a single simulation. The simulation set the lane above, which is the left-hand lane as a normal speed route, vehicles would move at one speed unit, and the other lane is for overtaking, where vehicles would have one speed unit higher. The intelligent agent would be able to change directions and speeds as discussed before. To simulate V2V communication, if the radar detects an obstacle vehicle through its environmental scanner, the agent will request for the speed of the vehicle. The obstacle vehicle would give its speed-reading to the agent, which enables the agent to use additional information to define the state of the environment.

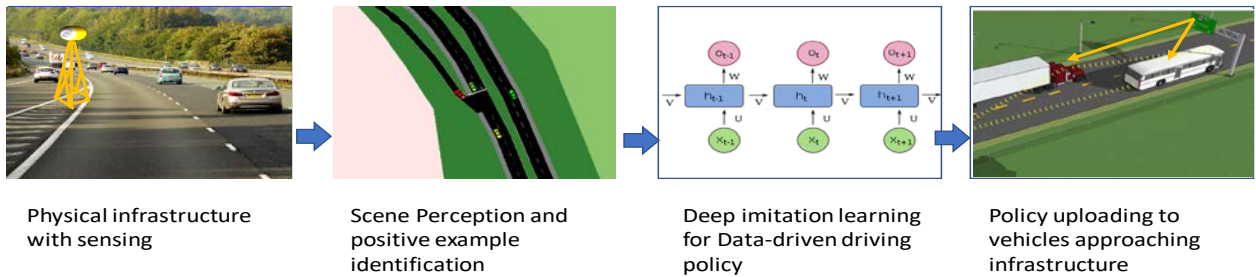

Fig. 3. System architecture for proposed Data-driven Vehicle-Infrastructure Policy Communication Scheme.

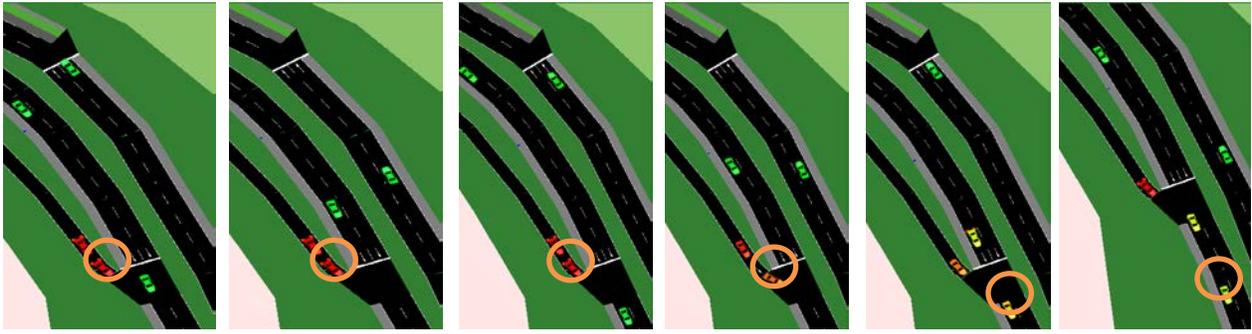

Fig. 4. An example sequence of states when a vehicle is trying to join a motorway/freeway.

## C. Deep Imitation Learning for Infrastructure-led Policy Learning

The above sections described how a CAV with sensing and V2V communication ability could be trained to learn behavior over time. While such a model is suitable for simple scenarios like straight line driving, CAV's often come to deal with more tricky scenarios such as negotiations at roundabouts, joining motorways (freeways) and giving way to pedestrians etc. Moreover, most of such negotiating situations could be unique in its appearance with a unique local context attached to it. E.g. there could be tricky points such as give-way situations, which may go unnoticed to the driver / CAV if it has not come across it before. Could the road-side infrastructure located at such points, look at the traffic scenarios and learn the optimal behavior of a vehicle that is approaching it? If this is possible, then, when a vehicle is approaching the tricky point of negotiation, the infrastructure could instruct the CAV of the best policy.

To handle such situations, in this section, we propose a novel algorithm for infrastructure-led policy learning. The system architecture for the proposed scheme for policy learning and communication is illustrated in Fig. 3. According to Fig. 3, the cameras and sensors based at key junctions and intersections will capture the passing vehicles and pedestrians. The captured data will be analyzed to recognize vehicle, pedestrians and other objects of interest. A sequential data classifier will classify positive negotiations from negative negotiations. The positive negotiations are temporal sequences of events without any casualties or collisions. Once positive sequences of events are identified, the event sequences are utilized to learn an optimal policy for vehicle control. The optimal policy is learnt through imitation learning. Recurrent neural networks provide a framework to efficiently learn patterns in sequential data streams. In particular, Long-Short Term Memory (LSTM) networks are suitable variation of recurrent neural networks that does not suffer from the vanishing gradient problem associated with vanilla recurrent neural networks.

An example sequence of a positive outcome is depicted in Fig. 4. The sequence of steps taken by the marked car is a successful sequence of actions that a driver could follow to join the motorway. Our proposal is that, similar outcomes can be collected and the sequential pattern can be learnt. To learn the sequential patterns, and to come up with an optimal policy, the state of the environment will be encoded in a vector that represents the distance and speed of each vehicle in its surrounding, and fed in to the LSTM network. The output of the LSTM network will be the speed and angle of the vehicle at each time step. The pattern learnt, will yield in an optimal policy. Now when a CAV that has not experience this setting in the past comes to negotiate at this junction, the learnt policy can be transferred to it. This way we can overcome the requirement for a CAV to experience situations multiple times to learn a policy.

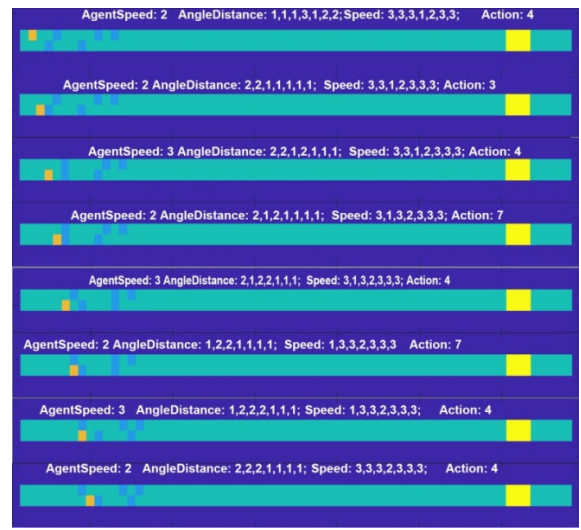

Fig. 5. A sequence sample of agent actions.

## D. Simulation Framework for Infrastructure-led Policy Learning

To illustrate the benefits of the proposed infrastructure-led policy learning scheme, we develop an experimental testbed based on a micro simulation framework and a python/Keras based neural network setup. While we do not have in possession a suitable data set, it is assumed that through sensing and perception an output similar to figure 4 can be obtained. Therefore, for the purpose of this simulation, the Simulation of Urban Mobility (SUMO) open source platform is utilized [19]. The output of the SUMO platform is dumped in XML format, which is processed to identify appropriate sequences of events, surrounding vehicles, and speed, and position of vehicles. Several positive outcomes are collected, and encoded, and fed to the LSTM network as training data. Once the network is trained, it will be used to predict the sequential actions of a CAV that is intending to join the motorway.

## IV. EXPERIMENTAL RESULTS AND DISCUSSIONS

This section describes the experimental testbed, the simulation results and the discussion of the results.

### A. Experimental Conditions

For evaluating dynamic programming-based policy learning method for CAV, a simulation environment has been implemented in Matlab 2017a environment. Three different experiments are performed to test the fundamental process of decision making. The learning rate α and discount factor γ is set to 0.4 and 0.95, respectively. The maximum speed of the agent is set at 3 units per time step.

The infrastructure-led policy learning scheme is simulated with SUMO micro-simulation platform (version 0.32.0) and the LSTM network is implemented based on the Keras/python package.

### B. Simulation Results

Fig. 5 shows a sequence of changes between agent and obstacle vehicles at different time steps. The orange square represents the location of agent; blue squares are the moving obstacle vehicles. The numbers above each picture are demonstrating the current speed of agent, seven readings from each direction, and the action it made to move it from previous states. From the initial five pictures, it shows that the agent tried to increase the speed and move to overtake other objects. From the last three pictures and changing speed values, it can be seen that a decision was taken to decrease the speed and follow the other vehicles, as there is no possible way to overtake them.

In the next set of results, we compare between the two situations where the agent learns with just the environment scanner and when the agent has access to speed information of the surrounding vehicles exchanged through V2V communication. Firstly, let's consider the average time cost, i.e. the number of time steps that the agent took to get from the starting point to its target successfully.

As the graph in Fig. 6(a) illustrates the time consumption to get to the destination gradually decreases as the number of iterations of experiment increases. This means that the agent self learns to make better decisions as it experiences more situations. It takes Approximately 25 units of time in average to reach the destination after 100,000 iterations of Q-learning. The pattern of the agent with vehicle communication, to share speed information of the surrounding neighbors, demonstrate to converge quickly than the scenario without the V2V communication. At the same time the time consumption to reach the destination gradually decreases too.

In Fig. 6(b), the probability of crashing illustrates the rate of agent having crashes with other vehicle or bumping to the wall. It clearly present that after a period of Q-learning, the agent would have the ability to avoid obstacles, while its moving and reduces the possibility of having crashes with others. Fig. 6(b) also presents that with V2V communication, the agent keeps a lower possibility of crashing, compared to agent without V2V communication, which has a higher potential to crash

Lastly, Fig. 6(c) shows the percentage of time an agent takes a faster route, which costs less than 40 units of time.

The initial results of the infrastructure-led policy learning framework can be illustrated as given in Fig. 7(a)-(c). Fig. 7(a) illustrates the LSTM network architecture that is trained for the scenario. The Fig. 7(b)-(c) illustrate two different instances of a vehicle trying to join a motorway. In the two different instances illustrated, we compare the variation of speed of the vehicle as given by the micro-simulation platform and the speed predicted by the proposed LSTM network at different time steps.

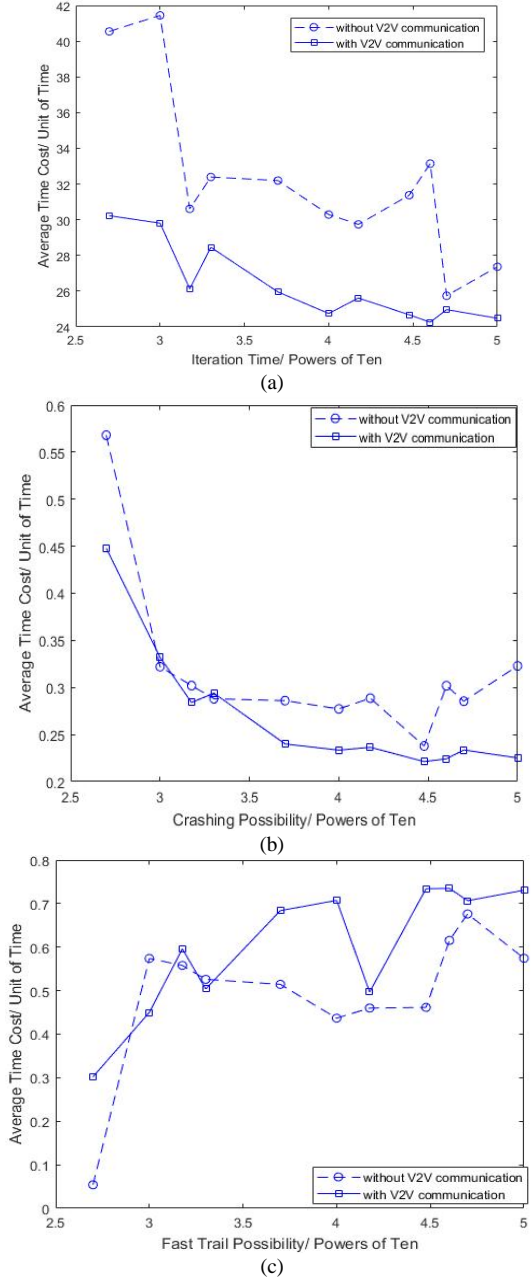

Fig. 6. Simulation results for comparison between the agent performance with and without V2V communication. (a) Average time to reach destination. (b) Amount of crashes (c) Frequency of quick finish (reach destination<40 time units).

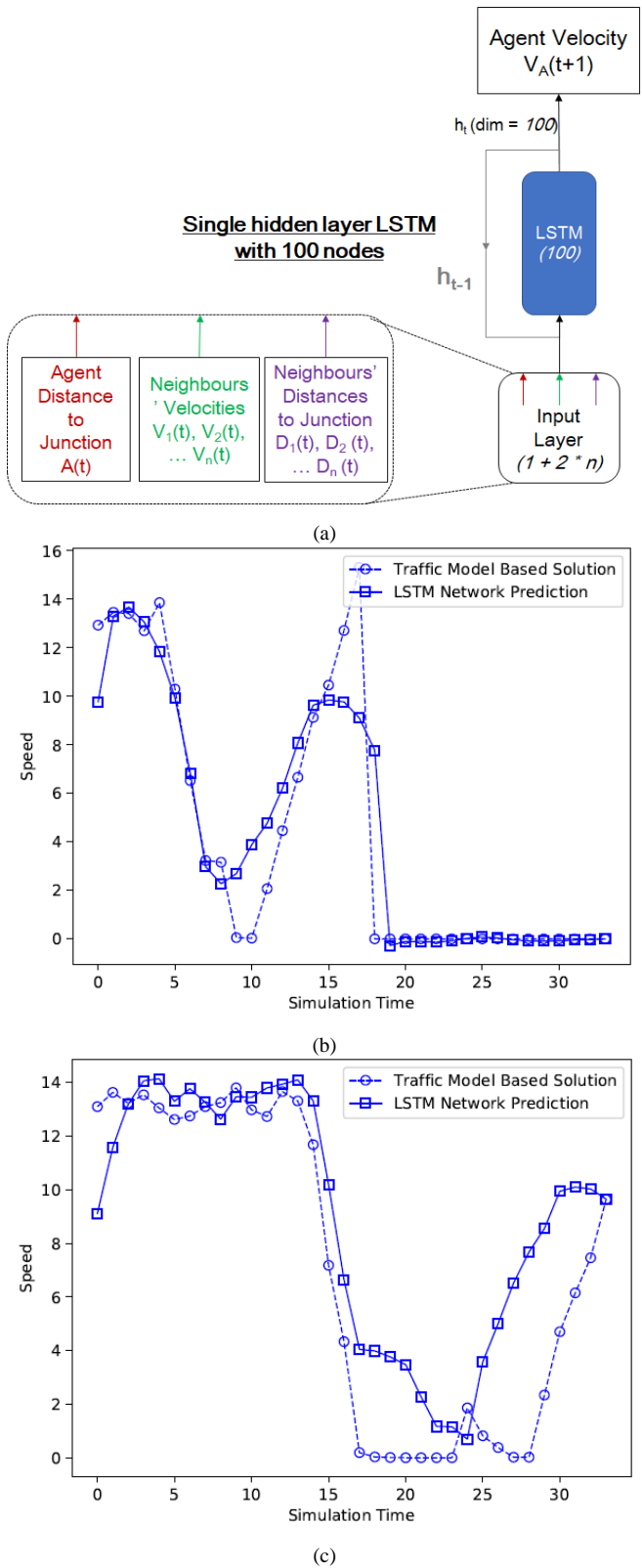

Fig. 7. Simulation results for infrastructure-led policy learning of a vehicle joining a motorway: (a) The LSTM network architecture and the variation of speed as approaching and joining the motorway. (b) test scenario 1. (c) test scenario 2.

*C. Discussion of Results*

The above results illustrate a significant improvement and an efficiency of learning after the first two thousand iterations. In the following iterations, the pattern of changing is slowing down. While the process of decision-making in the agent improves over the iterations, the agent with vehicle speed communication seems to have a better performance than the agent with only the environmental scanner.

In summary, agent with V2V communication would have a better efficiency in learning to make better driving policies. With the additional parameters obtained through V2V communication, the agent could perform better as it can explore of more states, due to unknown policies of the states it never reached before. Thus, the speed information shared by the vehicles acts as an important additional dimension which significantly improves the decision making capability of the simulated CAV.

Considering with the complexity of state locating, the agent with vehicle speed communication has approximately twice the number of dimensions to represent its state of the environment. As the simulation conditions become complex, such as for V2I communication or when considering real CAVs, this would lead to bottlenecks in terms of computational complexity. Deep reinforcement learning promises to tackle such situations of very high dimensions.

For the infrastructure-led policy learning framework, while the results illustrated in Fig. 7 are encouraging, the results could be improved further by tuning the structure of the LSTM network and increasing the number of training data sequences. It has to be noted, that only a simple LSTM network was setup in this case to prototype the idea.

V. CONCLUSIONS AND FUTURE WORK

The connected autonomous vehicles (CAVs) operate by sensing its surroundings and exchanging information from other vehicles and infrastructure. This paper investigated the possibility for CAVs to learn from these data sources and derive its own safe driving policies. A reinforcement learning framework is presented in the paper, which simulates the self-evolution of a CAV over its lifetime. The results indicated that overtime the CAVs are able to learn useful policies to avoid crashes and achieve its objectives in more efficient ways. Vehicle to vehicle communication enables additional useful information to be acquired by CAVs, which in turn enables CAVs to learn driving policies more efficiently. Furthermore, the paper proposed an infrastructure-led policy learning and communication framework based on deep imitation learning.

The initial validation results based on a microsimulation platform and a LSTM network were encouraging. The future work will investigate complex driving policies such as roundabout negotiations, cooperative learning between CAVs and deep reinforcement learning to traverse larger state spaces.

Furthermore, we will validate the infrastructure-led policy learning framework with real world datasets on vehicle tracking.

## REFERENCES

[1] T. Koslowski, "Your Connected Vehicle Is Arriving," MIT Technology Review. [Online]. Available: https://www.technologyreview.com/s/426523/your-connected-vehicle-is-arriving/. [Accessed: 31-Aug-2017].

[2] G. Pau, "Quickly Home Please: How Connected Vehicles Are Revolutionizing Road Transportation," IEEE Internet Computing, vol. 17, no. 1, pp. 80–83, Jan. 2013.

[3] N. Lu, N. Cheng, N. Zhang, X. Shen, and J. W. Mark, "Connected Vehicles: Solutions and Challenges," IEEE Internet of Things Journal, vol. 1, no. 4, pp. 289–299, Aug. 2014.

[4] R. Coppola and M. Morisio, "Connected Car: Technologies, Issues, Future Trends," ACM Comput. Surv., vol. 49, no. 3, p. 46:1–46:36, Oct. 2016.

[5] B. Reichelt, H. Ußler, O. Michler, J. Holfeld, S. Eckelmann, and T. Trautmann, "Model-based generation and validation of different sensor information contributing to a fusion algorithm in connected vehicles," in 2017 5th IEEE International Conference on Models and Technologies for Intelligent Transportation Systems (MT-ITS), 2017, pp. 339–344.

[6] J. Hu, L. Kong, W. Shu, and M.-Y. Wu, "Scheduling of connected autonomous vehicles on highway lanes," in 2012 IEEE Global Communications Conference (GLOBECOM), 2012, pp. 5556–5561.

[7] R. King, "Traffic management in a connected or autonomous vehicle environment," in Autonomous Passenger Vehicles, 2015, pp. 1–20.

[8] D. Jia, K. Lu, J. Wang, X. Zhang, and X. Shen, "A Survey on Platoon-Based Vehicular Cyber-Physical Systems," IEEE Communications Surveys Tutorials, vol. 18, no. 1, pp. 263–284, Firstquarter 2016.

[9] W. Gao, Z. P. Jiang, and K. Ozbay, "Data-Driven Adaptive Optimal Control of Connected Vehicles," IEEE Transactions on Intelligent Transportation Systems, vol. 18, no. 5, pp. 1122–1133, May 2017.

[10] Z. Huang, X. Xu, H. He, J. Tan, and Z. Sun, "Parameterized Batch Reinforcement Learning for Longitudinal Control of Autonomous Land Vehicles," IEEE Transactions on Systems, Man, and Cybernetics: Systems, pp. 1–12, 2017.

[11] P. Wang and C. Y. Chan, "Formulation of deep reinforcement learning architecture toward autonomous driving for on-ramp merge," in 2017 IEEE 20th International Conference on Intelligent Transportation Systems (ITSC), 2017, pp. 1–6.

[12] F. Liu, Y. Li, L. Zuo, and X. Xu, "Research on Lane Change Decision for Autonomous Vehicles Based on Multi-Kernels Least Squares Policy Iteration," in 2017 9th International Conference on Intelligent Human-Machine Systems and Cybernetics (IHMSC), 2017, vol. 1, pp. 385–389.

[13] X. Li, X. Xu, and L. Zuo, "Reinforcement learning based overtaking decision-making for highway autonomous driving," in 2015 Sixth International Conference on Intelligent Control and Information Processing (ICICIP), 2015, pp. 336–342.

[14] J. Wang, X. Xu, D. Liu, Z. Sun, and Q. Chen, "Self-Learning Cruise Control Using Kernel-Based Least Squares Policy Iteration," IEEE Transactions on Control Systems Technology, vol. 22, no. 3, pp. 1078–1087, May 2014.

[15] C. Innocenti, H. Lindén, G. Panahandeh, L. Svensson, and N. Mohammadiha, "Imitation learning for vision-based lane keeping assistance," in 2017 IEEE 20th International Conference on Intelligent Transportation Systems (ITSC), 2017, pp. 425–430.

[16] H. M. Le, P. Carr, Y. Yue, and P. Lucey, "Data-Driven Ghosting using Deep Imitation Learning," in MIT Sloan Sports Analytics Conference, 2017.

[17] "DeepMind's work in 2016: a round-up," DeepMind. [Online]. Available: https://deepmind.com/blog/deepmind-round-up-2016/. [Accessed: 01-Sep-2017].

[18] R. S. Sutton and A. G. Barto, Reinforcement Learning. MIT Press, 1998.

[19] "DLR - Institute of Transportation Systems - SUMO – Simulation of Urban MObility." [Online]. Available: http://www.dlr.de/ts/en/desktopdefault.aspx/tabid-9883/16931_read-41000/. [Accessed: 10-Apr-2018].